\documentclass[prb,amsmath,amssymb,12pt,draft,showpacs]{revtex4}
\usepackage{epsf}
\newcommand{\gae}{\hbox{\lower0.6ex\hbox{$\sim$}\llap{\raise0.6ex\hbox{$>$}}}}
\newcommand{\lae}{\hbox{\lower0.6ex\hbox{$\sim$}\llap{\raise0.6ex\hbox{$<$}}}}
\begin{document}
%\draft
\title{Surface and bulk transitions in three-dimensional O$(n)$ models}
\author{Youjin Deng~$^{1,2}$, Henk W.J. Bl\"ote~$^{2,3}$, 
and M. P. Nightingale~$^{4}$} 
\affiliation{$^{1}$Laboratory for Materials Science, Delft 
University of Technology,
Rotterdamseweg 137, 2628 AL Delft, The Netherlands}
\affiliation{$^{2}$Faculty of Applied Sciences, Delft University of
Technology, P.O. Box 5046, 2600 GA Delft, The Netherlands}
\affiliation{$^{3}$ Lorentz Institute, Leiden University,
  P.O. Box 9506, 2300 RA Leiden, The Netherlands}             
\affiliation{$^{4}$ Department of Physics, University of Rhode Island,
Kingston, Rhode Island 02881, USA} 
%\date{\today} 
\vskip -7mm
\begin{abstract} 
\vskip -9mm
Using Monte Carlo methods and finite-size scaling,
we investigate surface criticality in the O$(n)$ models
on the simple-cubic lattice
with $n=1$, $2$, and $3$, 
i.e. the Ising, $XY$, and Heisenberg models. 
For the critical couplings we find $K_{\rm c}(n=2)=0.454\,1655\,(10)$ 
and $K_{\rm c}(n=3)= 0.693\,002\,(2)$.
We simulate the three models with open surfaces
and determine the 
surface magnetic exponents at the ordinary transition 
to be $y_{h1}^{\rm (o)}=0.7374\,(15)$, $0.781\,(2)$, and 
$0.813\,(2)$ for $n=1$, $2$, and $3$, respectively.
Then we vary the surface coupling $K_1$ and locate the so-called
special transition at $\kappa_{\rm c} (n=1)=0.50214 \,(8)$
and $\kappa_{\rm c} (n=2)=0.6222\,(3)$, 
where $\kappa=K_1/K-1$. The corresponding surface thermal
and magnetic exponents are $y_{t1}^{\rm (s)} =0.715 \,(1)$ and 
$y_{h1}^{\rm (s)} =1.636\,(1)$ for the Ising model, and
$y_{t1}^{\rm (s)} =0.608 \,(4)$  and
$y_{h1}^{\rm (s)} =1.675 \,(1)$ for the $XY$ model. 
Finite-size corrections with an exponent close to $-1/2$ occur
for both models. Also for the Heisenberg model we find substantial
evidence for the existence of a special surface transition.
\end{abstract}
\maketitle 
\section{Introduction}
In the past decades, surface effects near a phase transition have been
investigated extensively, and many results have been obtained by means of
mean-field theory, series expansions, renormalization and field-theoretic
analyses. For reviews, see e.g. Refs.~\onlinecite{KB1,HWD1}, and for more
recent work see Refs.~\onlinecite{HWD5,MP1}. In particular, at a
second-order phase transition, where long-range correlations emerge,
surface effects can be significant. The surfaces display critical 
phenomena which differ from the bulk critical behavior; several 
surface universality classes can exist for one bulk universality 
class. We shall refer to the various types of transitions using
the terminology of Ref.~\onlinecite{KB1}.

\begin{figure}
\begin{center}
\leavevmode
\epsfxsize 10.0cm
\epsfbox{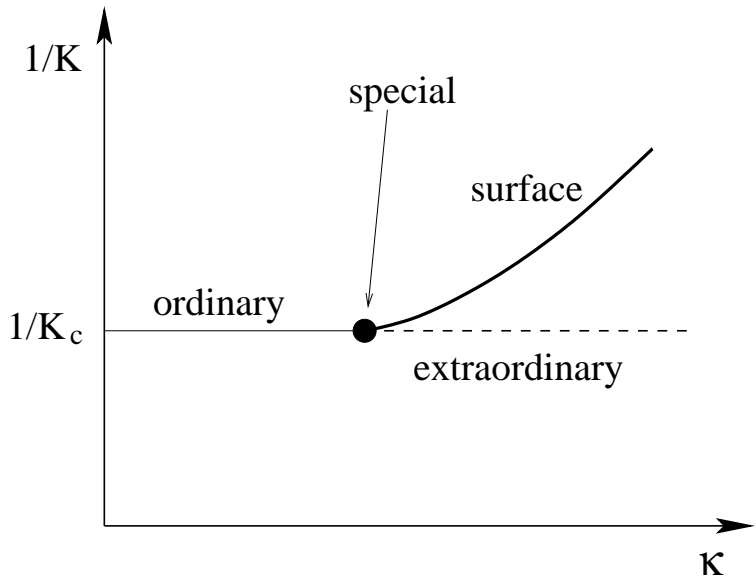}
\end{center}
\caption{Sketch of the surface phase transitions of the
three-dimensional Ising and $XY$ models with ferromagnetic
couplings.  The vertical axis is the bulk temperature $1/K$,
and the parameter $\kappa=(K_1-K)/K$ in
the horizontal axis represents the enhancement of the
surface couplings. The ``surface,'' the ``ordinary,'' and
the ``extraordinary'' phase transitions are represented by
the thick solid, the thin solid, and the dashed line, respectively.
The lines meet in a point, shown as the black circle, which
is referred to as the ``special'' phase transition.}
\label{fig01}
\end{figure}

In this work, we investigate surface critical phenomena in 
three-dimensional O$(n)$ models, namely the Ising $(n=1)$,
the $XY$ $(n=2)$, and the Heisenberg $(n=3)$ model.
The reduced Hamiltonian of these models
can be written as the sum of
two parts: a bulk term proportional to the volume of the system
and a surface term proportional to the surface area, i.e.,
\begin{equation}
{\cal H}/k_{\rm B}T =  - K {\sum_{\langle i j \rangle}}^{\rm (b)}
\vec{s}_{\rm i} \cdot \vec{s}_j -\vec{H} \cdot{\sum_k}^{\rm (b)} \vec{s}_k
- K_1{\sum_{\langle p q \rangle}}^{\rm (s)}\vec{s}_p\cdot \vec{s}_q -
\vec{H}_1 \cdot {\sum_r}^{\rm (s)} \vec{s}_r  \; ,
\label{Ham1}
\end{equation}
where the dynamic variable $\vec{s}$ is a unit vector of $n$ components.
The parameters $K$ and $K_1$ are the strenghts of the coupling between
nearest-neighbor sites in the bulk and on the surface layers, respectively,
and $H$ and $H_1$ represent the reduced magnetic fields.
The first two sums in  Eq.~(\ref{Ham1}) 
account for the bulk and the last two sums
involve the spins on the open surfaces. 
For ferromagnetic bulk and surface couplings ($K>0$ and $K_1>0$), 
the phase transitions are sketched in Fig.~\ref{fig01} for the case
of the Ising and the $XY$ model. In the high-temperature region,
i.e., for bulk coupling $K<K_{\rm c}$, the bulk is in the paramagnetic
state, so that the bulk correlation length remains finite.
However, a phase transition can still occur on the open surface
when the surface coupling $K_1$ is sufficiently enhanced.
This phase transition is referred to as the ``surface transition,''
and is represented by the solid curve in Fig.~\ref{fig01}.
These phase transitions are generally thought to be in the 
same universality classes as the two-dimensional 
Ising and the $XY$ model, respectively. At the bulk critical point
$K=K_{\rm c}$, the line of surface phase transitions terminates at a point
$(K_{\rm c}, K_{1{\rm c}})$. At this point, both the surface and the bulk 
correlation length diverge. 
Thus, the point $(K_{\rm c}, K_{1{\rm c}})$
acts as a multicritical point, and the phase transition is referred
to as the ``special transition.'' For $K_1<K_{1{\rm c}}$, the bulk and the 
surfaces simultaneously undergo a phase transition at $K=K_{\rm c}$. In 
this case, the critical correlations on the surfaces arise from 
the diverging bulk correlation lengths, and the transition is named
the ``ordinary transition.'' The ordinary transition remains within
the same universality class  for a wide range of surface couplings.
The correlation functions on and near the surface appear to fit 
universal profiles\cite{MPN1}. The transitions at 
$K=K_{\rm c}$ for $K_1>K_{1{\rm c}}$ are referred to as the ``extraordinary 
transitions.'' For the Ising model, since the surfaces 
are already in the ferromagnetic
state for a smaller coupling $K<K_{\rm c}$, no surface transition 
occurs when the bulk critical line $K=K_{\rm c}$ is crossed. Nevertheless,
owing to the diverging bulk correlation length, the surfaces still
display critical correlations at $K=K_{\rm c}$. 
For the $XY$ model, however, the surface transitions 
for $K<K_{\rm c}$ are Kosterlitz-Thouless-like~\cite{JMK},
i.e., the surfaces do not display long-range order for $K<K_{\rm c}$,
in agreement with results of Landau and co-workers~\cite{DPL1}.

For three-dimensional O$(n)$ models with $n>2$, which include the
Heisenberg model, the line of surface transitions for $K<K_{\rm c}$ does not
exist; it may thus seem self-evident that the special and the extraordinary
transitions are also absent. However, this remains to be investigated;
for instance, in two-dimensional tricritical Potts models, a line of
edge transitions is absent, but special and extraordinary transitions
do exist \cite{YD6}.
Thus, even without a line of surface transitions for $K<K_{\rm c}$, rich
surface critical phenomena can still occur in the three-dimensional
Heisenberg model. For instance, it was reported~\cite{MK} that at bulk
criticality $K=K_{\rm c}$ the surface magnetic exponents depend on the
ratio $K_1/K$ for $K_1/K \geq 2.0$. This brings up the question whether

Additional surface critical phenomena can occur for the Ising model,
if the surface and/or the bulk couplings are allowed to
be antiferromagnetic.
Further, one can allow the spins on the surface to vanish,
such that the surface part of the Hamiltonian in Eq.~(\ref{Ham1}) is
described by the so-called Blume-Capel model. Such spin-0 states act
as annealed vacancies on the surfaces. It was observed~\cite{YD3} that, 
by varying the fugacity of the vacancies, one can reach a point where 
the bulk Ising criticality $K=K_{\rm c}$ joins the line of surface
transitions that belongs to the universality class
of the two-dimensional tricritical Ising model. This point was 
named~\cite{YD3} the ``tricritical special'' phase transition.
In short, for each bulk universality class, surface transitions in
various surface universality classes can occur, including the
ordinary, special, and extraordinary transitions at $K=K_{\rm c}$,
and the surface transitions at $K<K_{\rm c}$. 

Apart from 
the bulk renormalization exponents, additional surface exponents are 
needed to describe  the above surface critical behavior. 
At the ordinary and the extraordinary transitions, the surface
magnetic scaling field is relevant, while the surface thermal field
is irrelevant. At the special transition, both the magnetic and the
thermal surface fields are relevant. 

Since exact information about
critical behavior is scarce in three dimensions, determinations of these
surface critical exponents
rely on approximations of various kinds. These 
include mean-field theory~\cite{KB1,KB2,KB3,TCL},
series expansions~\cite{KO}, renormalization group 
technique~\cite{HWD1,HWD5,HWD2,HWD3,HWD4}, Monte Carlo
simulations~\cite{MPN1,DPL1,DPL2,CR2,CR1,MP2,KB4}, etc.

The surface critical index $\beta_1$ is defined so as to describe the
asymptotic scaling behavior of the surface magnetization $m_1$ as a
function of the bulk thermal field $t$, i.e., $m_1 \propto t^{\beta_1}$.
From the scaling relations it follows that this exponent is related
to the critical exponents as $\beta_1=(d-1-y_{h1})/y_t$, where $y_t$ and
$y_{h1}$ are the bulk thermal and the surface magnetic exponent,
respectively, and $d=3$ is the spatial dimensionality.
The mean-field analysis and the Gaussian fixed point
of the $\phi^4$ theory yield the magnetic surface
index $\beta_1$ as $\beta_1^{\rm (o)}=1$, $\beta_1^{\rm (s)}=1/2$, and
$\beta_1^{\rm (e)}=1$ respectively for the ordinary, special, and
extraordinary transition. 
Many numerical results also exist. For the simple-cubic lattice, the 
special transition of the Ising model was located as
$\kappa_{\rm c}=0.5004 \,(2)$~\cite{CR2,CR1}. Although the values of
critical couplings $K_{\rm c}$ and $K_{1{\rm c}}$ are far from the
mean-field predictions, 
the above result for $\kappa_{\rm c}$ is in agreement
with the mean-field value $\kappa_{\rm c}=1/2$. Further, 
the surface critical exponents
are determined~\cite{CR2,CR1,MP2,YD1} 
as $y_{h1}^{\rm (o)}=0.737 \,(5)$, $y_{h1}^{\rm (s)}=1.62 \,(2)$,
and $y_{t1}^{\rm (s)}=0.94 \,(6)$.   Compared to the Ising model,
there are fewer investigations for the three-dimensional $XY$ and the
Heisenberg model. 
In particular, to our knowledge,
numerical determinations of the special transition and the
corresponding surface critical exponents have not yet been reported
for the $XY$ model. Most of the existing results for the Ising, the
$XY$ and the Heisenberg model will be tabulated below, together
with new results of the present work.

The present work aims to provide an extensive and systematic Monte
Carlo investigation of the phase transitions of the surfaces of
the three-dimensional Ising, $XY$, and Heisenberg models.
Compared to numerical investigations one or two
decades ago, one has the following advantages. 
Firstly, the bulk critical points of these systems have now been determined
accurately. On the simple cubic lattice, the bulk critical
point of the Ising model  was determined as
$K_{\rm c}(n=1)=0.221\,654\,55 \,(3)$~\cite{YD2}, with the uncertainty only
in the eighth decimal place. The bulk transitions of the $XY$ and the 
Heisenberg model were also determined~\cite{KO,MPN2,MPN3,JA,Ad,AC,HGB} to
occur at $K_{\rm c}=0.454\,167 \,(4)$ and $0.693\,002 \,(12)$, respectively.
In the present paper, we also simulate these two systems with 
periodic boundary conditions, and improve the above estimations 
as $K_{\rm c}(n=2)=0.454\, 1655\,(10)$ and $K_{\rm c}(n=3)=0.693\, 002 \,(2)$.
Secondly, the rapid development of computer technology makes it
possible to perform extensive computations at a limited cost.
The present work was performed on 20 PCs; the total computer time 
is in the order of 20 CPU months at a processor speed of  2.5 GHz.

The organization of the present paper is as follows.  Section II
reviews the finite-size-scaling properties of the systems defined
by Eq.~(\ref{Ham1}), with the emphasis on the sampled quantities 
required for the numerical analysis of the simulation data.
Section III describes the determination of the critical points of
the $XY$ and Heisenberg models. Sections IV, V, and VI present the Monte
Carlo simulations and the results for the Ising, $XY$, and Heisenberg
models, respectively. Section VII concludes the paper with a brief discussion.

\section{Finite-size scaling and sampled quantities}
The total free energy of a system with free surfaces can,
in analogy with the Hamiltonian in Eq.~(\ref{Ham1}), be expressed as the sum
of a bulk and a surface term~\cite{KB1,MEF1,GC}:
\begin{equation}
F=f_{\rm b} V +f_1 S \; ,
\end{equation}
where $f_{\rm b}$ and $f_1$ are the densities of the bulk and the surface 
parts of the free energy, respectively, and $V$ and $S$ represent
the total volume and the surface area, respectively. Near criticality,
the finite-size scaling behavior of $f_{\rm b}$ and $f_1$ is given by
the equations 
\begin{equation}
f_{\rm b}(t, h,  L) = L^{-d} f_{\rm bs}(t L^{y_t},
h  L^{y_h})+f_{\rm ba}(t, h) \; ,
\label{bfree}
\end{equation}
and
\begin{equation}
f_1(t, h, t_1, h_1, L) = L^{-(d-1)} f_{\rm 1s}(t L^{y_t}, h  L^{y_h},
t_1 L^{y_{t1}}, h_1  L^{y_{h1}}) +f_{\rm 1a}(t, h, t_1, h_1) \; .
\label{sfree}
\end{equation}
The functions $f_{\rm bs}$ and $f_{\rm ba}$ are the singular and
the analytical parts of $f_{\rm b}$; $f_{\rm 1s}$ and $f_{\rm 1a}$
similarly apply to the surface free-energy density $f_1$.
The bulk thermal and magnetic scaling fields are represented
by $t$ and $h$, and the surface scaling fields by $t_1$ and $h_1$. 
The associated exponents are labeled with corresponding subscripts.
As implied by Eq.~(\ref{bfree}), the leading scaling behavior of
the bulk does not depend on the presence of free surfaces, although
physical quantities near the surfaces can be significantly affected. 

On the basis of Eqs.~(\ref{bfree}) and (\ref{sfree}), the scaling behavior
of various quantities can be obtained as derivatives of $f_{\rm b}$ and
$f_{\rm s}$ with respect to the appropriate scaling fields. Details can
be found in Ref.~\onlinecite{KB1}.

The determination of the bulk critical points used simulations of
$L \times L \times L$ with periodic boundary conditions in which case
$f_1$ vanishes. The sampling procedure involved the determination
of the bulk magnetization density
\begin{equation}
\vec{m}\equiv N^{-1} \sum_{k=1}^N \vec{s}_k,
\label{m2}
\end{equation}
where $N=L^{3}$. This yielded the averages of the
magnetization moments $\langle \vec{m} \cdot \vec{m} \rangle$ and
$\langle (\vec{m} \cdot \vec{m})^2 \rangle$.
The quantity
\begin{equation}
Q(K,L) \equiv \frac{\langle \vec{m} \cdot \vec{m} \rangle^2}
{\langle (\vec{m} \cdot \vec{m})^2 \rangle}\,,
\label{Q}
\end{equation}
which is related to the Binder cumulant~\cite{KB}, converges to
a universal value $Q$ at the critical point, and was used to
determine the critical coupling $K_{\rm c}$. The finite-size scaling
behavior of $Q$ can be found by writing the moments of $\vec{m}$
in terms of derivatives of the free energy with respect to the magnetic field.
After application of a scaling transformation, the singular powers
in $Q$ associated with field derivatives cancel,
as do the powers of the nonuniversal metric factor relating the physical
field and the magnetic scaling field.
In the vicinity of the critical point one
obtains, in terms of the temperature scaling field $t$ and an
irrelevant temperature-like field $u$,
\begin{equation}
\label{q00}
Q(t,u,L)=\tilde{Q}(tL^{y_t},uL^{y_{\rm i}})+b_2L^{3-2y_h}+b_3L^{y_t-2y_h}
+\cdots
\end{equation}
where $y_{\rm i}$ is the leading irrelevant
exponent. The correction term with amplitude $b_2$ is due to the
analytic contribution to the second moment of $\vec{m}$, and that with
amplitude $b_3$ to the second-order dependence of the temperature
field on the physical magnetic field. Apart from corrections, the
temperature field is proportional to $K-K_{\rm c}$.
Eq.~(\ref{q00}) will be used in Sec. \ref{sec.crit} to
determine the bulk critical points.

In order to investigate surface critical behavior, we simulated
$L \times L \times L$ simple-cubic lattices with periodic
boundary conditions in the $xy$ plane and free boundaries in the $z$
direction.  First, we sampled the components of the surface
magnetization and obtained two generalized surface susceptibilities:
\begin{equation}
\chi_{11}= \frac{L^{2}}{2} \langle \vec{m}_1 \cdot \vec{m}_1 +
\vec{m}_2 \cdot \vec{m}_2\rangle \; ,
\hspace{5mm} \mbox{and} \hspace{5mm}
\chi_{12}= L^{2} \langle \vec{m}_1 \cdot \vec{m}_2 \rangle
\label{susc0}
\end{equation}
where $\vec{m}_1$ and $\vec{m}_2$ are the magnetization densities
at the free surfaces with $z=1$ and $z=L$, respectively.
By differentiating the surface free energy with respect
to magnetic fields that act on either one of the free surfaces, one
finds that the singular parts of these surface susceptibilities
scale as $L^{2y_{h1}-2}$.

In addition, we computed two surface-surface correlations
\begin{equation}
g_{11}=\frac{1}{2L^2} \sum_{x,y=1}^L (\langle \vec{s}_{x,y,1} \cdot
\vec{s}_{x+r,y+r,1} +\vec{s}_{x,y,L} \cdot \vec{s}_{x+r,y+r,L})
 \rangle \hspace{5mm} (r=L/2) \; ,
\label{cor0}
\end{equation}
and
\begin{equation}
g_{12}=\frac{1}{L^2} \sum_{x,y=1}^L \langle \vec{s}_{x,y,1} \cdot
\vec{s}_{x,y,L} \rangle \; .
\label{cor1}
\end{equation}
Further, we sampled two ratios of surface magnetization moments:
\begin{equation}
Q_{11}=\frac{ \langle \vec{m}_1 \cdot \vec{m}_1 \rangle^2}{
\langle (\vec{m}_1 \cdot \vec{m}_1)^2 \rangle}
\hspace{5mm} \mbox{and} \hspace{5mm}
Q_{12}=\frac{ \langle \vec{m}_1 \cdot \vec{m}_2 \rangle^2}{
\langle (\vec{m}_1 \cdot \vec{m}_2)^2 \rangle} \, .
\label{bind0}
\end{equation}
These quantities are the surface analogs of the bulk ratio $Q$,
cf.~Eq.~(\ref{q00}),
and will be used to locate the surface transitions.

\section{Critical points of the O(2) and the O(3) models}
\label{sec.crit}
\begin{table}
\caption{Description of the simulations of the {\em XY} and
Heisenberg models. The table lists the simulation length in
millions of cycles (\#MC), and the number of Wolff clusters (\#Wc/C)
per cycle, for each system size $L$.
The data were taken range $\Delta K$ of the coupling $K$. The values
shown are those for the {\em XY} model; those for the Heisenberg model
are approximately the same.\\}
\label{tab:simul}
\begin{tabular}{||r|r|r|l||}
\hline
$L$   & \#MC  & \#Wc/C & $\Delta K$\\
\hline
  4   &    50 &      2 &   0         \\
  6   &    50 &      3 &   0.016     \\
  8   &    50 &      4 &   0         \\
 10   &    20 &      5 &   0         \\
 12   &    20 &      6 &   0         \\
 14   &    20 &      7 &   0         \\
 16   &    80 &      8 &   0.006     \\
 20   &    20 &     10 &   0         \\
 24   &    20 &     12 &   0         \\
 28   &    20 &     14 &   0         \\
 32   &    80 &     16 &   0.002     \\
 40   &    20 &     20 &   0         \\
 48   &    20 &     24 &   0         \\
 64   &    20 &     32 &   0         \\
 96   &    15 &     48 &   0         \\
160   &   6.7 &     80 &   0         \\
\hline
\end{tabular}
\end{table}
The critical point of the Ising model on the simple cubic lattice is
already known \cite{YD2} with
sufficient accuracy for the present purposes.
We therefore restrict ourselves to the $XY$ and Heisenberg
models. We used the Wolff cluster algorithm~\cite{UW1,UW2} to
simulate these models on simple-cubic lattices with periodic boundary
conditions. Each cluster is constructed on the basis of one component
of the spin vectors. The cluster formation process is thus very similar
to that for the Ising model. Each simulation consists of a large number
of cycles, each of which contains several Wolff steps and a data
sampling procedure. The Wolff cluster flips do not change the absolute
values of the spin components. Thus, to satisfy ergodicity, each cycle
also includes a random rotation of the whole system of spin vectors.
We simulated a number of $L^3$ systems whose finite sizes $L$ 
are listed in Table \ref{tab:simul}, together with the number of
Wolff clusters per cycle and the total number of cycles per system size.

Most simulations of the $XY$ model took place at $K= 0.454\,15$,
and of the Heisenberg model at $K=0.693$. Both values are already very
close to the final estimates that we shall report for the respective
critical points. 
To avoid bias effects
associated with short binary shift registers \cite{disp,DISP}
we took two such shift registers, with lengths equal to
the Mersenne exponents 127 and 9689, and added the resulting two
maximum-length bit sequences modulo 2. This procedure leads to a
sequence whose leading deviation from randomness is a 9-bit
correlation, which is a considerable improvement in comparison
with the usual 3-bit correlations \cite{SB}.

The simulations yielded data for the Binder cumulant as described in the
preceding Section. 
Expanding $\tilde{Q}$ in Eq.~(\ref{q00}) and expressing the temperature
deviation from the critical point in
$K-K_{\rm c}$\,, leads to
\begin{equation}
\label{q01}
Q(K, L) = Q + a_1 (K-K_{\rm c})L^{y_t} +a_2 (K-K_{\rm c})^2 L^{2y_t} +
\cdots + b_1 L^{y_{\rm i}} +b_2L^{3-2y_h}+b_3L^{y_t-2y_h} + \cdots
\end{equation}
where $Q$ is a universal constant and the correction term with amplitude
$b_1$ is due to the irrelevant field. This expression was used to analyze
the numerical data for $Q(K, L)$ by means of least-squares fits.
The exponents were set to the estimates obtained by Guida and
Zinn-Justin \cite{GZJ}, namely
$y_t=1.492$, $y_{\rm i}=-0.789$ and $y_h=2.482$ for the $XY$ model, and
$y_t=1.414$, $y_{\rm i}=-0.782$ and $y_h=2.482$ for the Heisenberg model.
In order to determine the amplitudes $a_1$ and $a_2$ we included
some data for relatively small ($L= 8,16$ and 32) systems, taken at
values of $K$ differing up to the order of one percent from $K_{\rm c}$.
Satisfactory fits, as judged from the residual $\chi^2$ compared
with the number of degrees of freedom, were obtained including all
system sizes down to $L=4$ for the $XY$ and $L=6$ for the Heisenberg
model. We found that mixed terms proportional to
$(K-K_{\rm c})L^{y_{\rm i}+y_t}$ were insignificant.

\begin{table}
\caption{Summary of recent results for the critical coupling $K_{\rm c}$
of the three-dimensional $XY$ and Heisenberg models on the 
simple-cubic lattice with nearest-neighbor interactions. The error
margin in the last decimal place is shown in parentheses.}
\label{tab:compare}
\begin{center}
\begin{tabular}{||l|l|c|lr||}
\hline
Reference                        & model & Year & $K_{\rm c}$      &     \\
\hline
Janke \cite{WJ}                  & O$(2)$  & 1993 & $0.454\,08$  & (8) \\
Adler et al.~\cite{Ad}           & O$(2)$  & 1993 & $0.454\,14$  & (7) \\
Gottlob and Hasenbusch \cite{GH} & O$(2)$  & 1993 & $0.454\,20$  & (2) \\
Butera and Comi \cite{BC}        & O$(2)$  & 1997 & $0.454\,19$  & (3) \\
Ballesteros et. al~\cite{HGB}    & O$(2)$  & 1996 & $0.454\,165$ & (4) \\
Cucchieri et. al~\cite{AC}       & O$(2)$  & 2002 & $0.454\,167$ & (4) \\
Present work                     & O$(2)$  & 2004 & $0.454\,1655$&(10) \\
Chen et al.~\cite{CFL}           & O$(3)$  & 1993 & $0.693\,035$ &(37) \\
Holm and Janke  \cite{HJ}        & O$(3)$  & 1993 & $0.693\,0$   & (1) \\
Butera and Comi \cite{BC}        & O$(3)$  & 1997 & $0.693\,05$  & (4) \\
Ballesteros et al.~\cite{HGB}    & O$(3)$  & 1996 & $0.693\,002$ &(12) \\
Present work                     & O$(3)$  & 2004 & $0.693\,002$ & (2) \\
\hline
\end{tabular}
\end{center}
\end{table}
The results for the critical points are $K_{\rm c}=0.454\,1655\,(10)$ for
the $XY$ model and $K_{\rm c}=0.693\,002\,(2)$ for the Heisenberg model. 
The universal values of the amplitude ratios are $Q=0.8049\,(2)$
for the $XY$ model and $Q=0.8776\,(2)$ for the Heisenberg model.
We obtained similar results with other types of fits, which involved
fewer correction terms and excluded some  of the smallest system
sizes so as to obtain satisfactory residuals. This internal
consistency confirms that our error estimates are realistic.
The present results and some recent values taken from the literature
are summarized in Table \ref{tab:compare}.

\section{Ising model}
Although the three-dimensional Ising model has not been exactly solved, 
considerable information about its critical behavior is available 
from extensive investigations using various kinds of
approximations. For a review see, e.g., Ref.~\cite{KB6}.
For instance, evidence has been found that the Ising model is
conformally invariant in three dimensions~\cite{YD1,YD4}.
There is some consensus that the values of the bulk 
thermal and magnetic exponents, $y_t$ and $y_h$, are $1.587$ and 
$2.482$, respectively, with uncertainty only in the last decimal 
place. The bulk critical points of a variety of three-dimensional
systems with Ising universality have also been obtained \cite{YD2};
the bulk transition of the Ising model with nearest-neighbor
interactions on the simple-cubic lattice was determined as
$K_{\rm c}=0.221\,654\,55 \,(3)$.
The present work conveniently chooses this model so that no
further work to determine $K_{\rm c}$ is required. As mentioned
earlier, periodic boundary conditions are imposed in the
$xy$ plane and free boundaries along the $z$ direction.

\subsection{Ordinary phase transition}
Using the Wolff cluster algorithm~\cite{UW1,UW2}, we simulated the
Ising model at bulk criticality, with the surface couplings 
chosen equal to the bulk couplings, i.e., $K_1=K=K_{\rm c}$.
The system sizes were taken as $16$ 
even values in the range $4 \leq L \leq 48$.
During the Monte Carlo simulations, we sampled the surface 
susceptibilities $\chi_{11}$ and $\chi_{12}$, and the correlation 
functions $g_{11}$ and $g_{12}$.
To estimate $y_{h1}^{\rm (o)},$ the universal surface magnetic
exponent of the ordinary surface transition, we modeled the Monte
Carlo data for the surface susceptibilities $\chi_{11}$ and
$\chi_{12}$ by expressions of the form
\begin{equation}
\chi_1 (L) =\chi_{\rm a}+L^{2y_{h1}^{\rm (o)}-2} (b_0+b_1 L^{y_{\rm i}}
+b_2 L^{y_{t1}^{\rm (o)}} +b_3 L^{y_3} +b_4 L^{y_4}) \; ,
\label{fitchi0}
\end{equation}
where $\chi_{\rm a}$ and the $b_{\rm i}$ are non-universal and depend on the
characteristics of the surface; $\chi_1$ stands for either one of
$\chi_{11}$ and $\chi_{21}$.  The various parameters in this
expression were determined by a least-squares fit.  We set $\chi_{\rm a}=0$
to fit $\chi_{12}$.

Similarly, we fitted data for the correlation functions $g_{11}$ and
$g_{12}$ to expressions of the form
\begin{equation}
g_1 (L) =L^{2y_{h1}^{\rm (o)}-4} (b_0+b_1 L^{y_{\rm i}}
+b_2 L^{y_{t1}^{\rm (o)}} +b_3 L^{y_3} +b_4 L^{y_4}) \; ,
\label{fitcor0}
\end{equation}
Again, $g_1$ can be either $g_{11}$ or $g_{12}$; the non-universal
amplitudes $b_{\rm i}$ are fitting parameters independent of the
corresponding amplitudes in Eq.~(\ref{fitchi0}), although we use the
same symbols.

The correction terms with amplitudes $b_1$, $b_2$, $b_3$, and $b_4$ in
Eqs. (\ref{fitchi0}) and (\ref{fitcor0}) account for the leading
finite-size corrections.  The exponent $y_{\rm i}=-0.821 \,(5)$~\cite{YD2} is
the leading irrelevant thermal scaling field in the three-dimensional
Ising universality class.  Further, since the thermal surface scaling
field for the ordinary transition is irrelevant, it may also introduce
finite-size corrections.  From a simple scaling argument it can be
derived that the value of this irrelevant surface exponent is
$y_{t1}^{\rm (o)}=-1$~\cite{TWB}, independent of the spatial
dimensionality. In principle, finite-size corrections from other
sources can occur, so that we also include the terms with amplitudes
$b_3$ and $b_4$.  We simply took $y_3=-2$ and $y_4=-3$.

Separate fits of the $\chi_{11}$ and $\chi_{12}$ data, employing
Eq.~(\ref{fitchi0}), yield consistent estimates:
$y_{h1}^{\rm (o)}=0.736 \,(2)$ and $0.738 \,(2)$, respectively. 

Fits of $g_{11}$ and $g_{12}$ yield $y_{h1}^{\rm (o)}=0.737 \,(2)$
and $0.736 \,(2)$, respectively.
A joint fit of both sets of susceptibility data, as well as one of
both sets of correlation function data, employing a single parameter
$y_{h1}^{\rm (o)}$ and independently variable amplitudes, yielded
consistent results but no significant improvement of the accuracy. 

We also simulated Ising systems in which the surface enhancement
is defined as in Ref.~\onlinecite{MPN1}. These systems differ 
from Eq.~(\ref{Ham1}) as to the couplings between the
surface layer and the second layer. We thus introduce an
enhancement parameter $\epsilon$ and define
couplings $K_1=\epsilon^2 K $ between nearest-neighbor sites
on the surface, and couplings $K_1'=\epsilon K $ between surface
sites and their nearest neighbors in next layer.  Whenever we 
parametrize the surface enhancement by $\epsilon$ we refer to
the Hamiltonian defined in Ref.~\onlinecite{MPN1}, which differs
from Eq.~(\ref{Ham1}).

\begin{figure}
\begin{center}
\leavevmode
\epsfxsize 10.0cm
\epsfbox{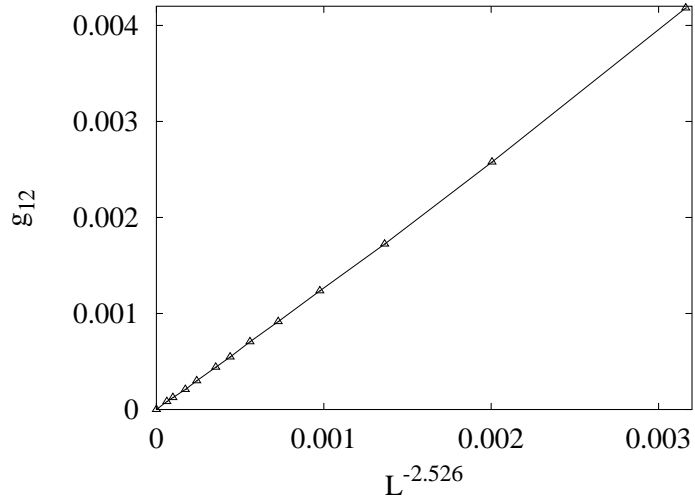}
\end{center}
\caption{Surface correlation function $g_{12}$ vs. $L^{-2.526}$ for
the Ising model with $\epsilon=0.8$. The error margins, in this figure
as well as in the following ones, are of the same order as the
thickness of the lines.}
\label{fig02}
\end{figure}
By varying the parameter $\epsilon$, one can move closer to the fixed
point for the ordinary phase transition so as to reduce the amplitudes
of finite-size corrections.  Systems with $\epsilon=1$ reduce to those
described above.  In accordance with Ref.~\onlinecite{MPN1}, in the
present work we also chose $\epsilon=0.9$ and $0.8$.  The analyses of the
data for the surface susceptibilities and the correlation functions
again employ Eqs.~(\ref{fitchi0}) and (\ref{fitcor0}); the results for
the surface magnetic exponents are in agreement with those obtained
for the case $\epsilon=1$.  As an illustration, the data for $g_{12}$
with $\epsilon=0.8$ are shown versus $L^{2y_{h1}^{\rm (o)}-4}$ in
Fig.~\ref{fig02}, where $y_{h1}^{\rm (o)}=0.737$ is taken from the
fit.

\begin{table}
\caption{Summary of the results for the surface critical exponents
in the three-dimensional Ising model, as obtained by different techniques.
MF: mean-field theory, MC: Monte Carlo simulations, FT: field-theoretical
methods, CI: Conformal invariance. The MF 
values of $y_{t1}$ and $y_{h1}$ have already 
made use of the mean-field predictions for the bulk thermal and 
magnetic exponents, which are $y_t=3/2$ and $y_h=9/4$, respectively.}
\label{tab_1}
\begin{center}
\begin{tabular}{||l|ll|llll||}
             & \multicolumn{2}{c|}{\em ordinary}   
             & \multicolumn{4}{c||}{\em special}  \\
\hline
             &$y_{h1}$            &$\beta_1$  
&$y_{h1}$    &$y_{t1}$            &$\beta_1$       &$\phi$ \\
\hline
MF~\cite{KB1,KB2}  &$1/2      $      &$ 1        $
&$5/4 $            &$3/4      $      &$1/2       $     &$1/2 $      \\
MC~\cite{DPL2}     &$0.72 \,(3) $    &$0.78 \,(2)  $
&$1.71 \,(16)$     &$0.94 \,(5) $    &$0.18 \,(2)  $   &$0.59\,(4)$  \\
MC~\cite{CR1}      &$0.721\,(6) $    &$0.807\,(4)  $
&$1.623\,(3 )$     &$ --      $      &$0.2375\,(15)$   &$ --    $    \\
MC~\cite{MPN1}     &$0.740\,(15)$    &$ --       $
&$ --      $       &$  --     $      &$ --       $     &$ --    $    \\
MC~\cite{MP2}      &$0.73 \,(1) $    &$0.80 \,(1)  $
&$ --      $       &$  --     $      &$ --       $     &$ --    $    \\
MC$+$CI~\cite{YD1} &$0.737\,(5) $    &$0.798\,(5)$
&$ --      $       &$  --     $      &$ --       $     &$ --    $    \\
MC~\cite{CR2}      &$ --      $      &$ --       $
&$1.624\,(8) $     &$0.73 \,(2) $    &$0.237 \,(5) $   &$0.461\,(15)$\\
FT~\cite{HWD1,HWD2}&$0.737    $      &$0.796     $
&$1.583    $       &$0.855    $      &$0.263     $     &$0.539    $  \\
FT~\cite{HWD3}     &$0.706    $      &$0.816     $
&$--       $       &$--       $      &$--        $     &$--       $  \\
FT~\cite{HWD4}     &$--       $      &$--        $
&$1.611    $       &$1.08     $      &$0.245     $     &$0.68     $  \\
\hline
Present            &$0.7374\,(15)$   &$0.796\,(1)$
&$1.636\,(1) $     &$0.715\,(1)$     &$0.229\,(1)  $   &$0.451\,(1) $ 
\end{tabular}
\end{center}
\end{table}
Finally, a joint fit to the data for $\chi_{11}$ and $\chi_{12}$ for
the three cases $\epsilon=1.0$, $0.9$, and $0.8$ yields $y_{h1}^{\rm
  (o)}=0.7374 \,(15)$; this result is in good agreement with most of the
existing results, as shown in Table~\ref{tab_1}.

\subsection{Special phase transition}
\begin{figure}
\begin{center}
\leavevmode
\epsfxsize 10.0cm
\epsfbox{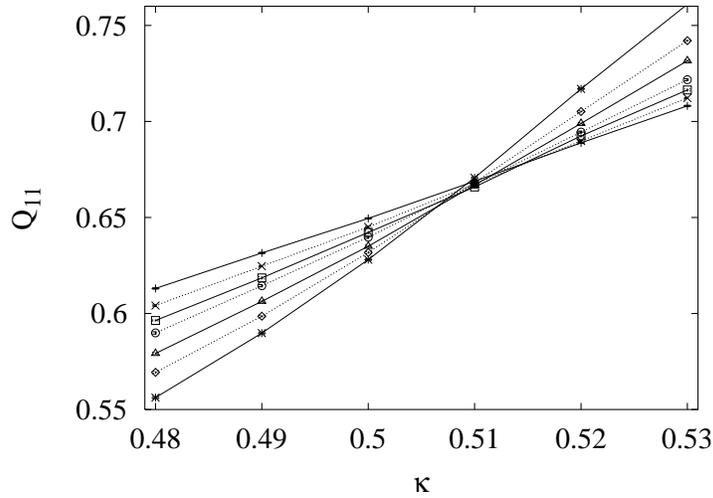}
\end{center}
\caption{Surface dimensionless ratio $Q_{11}$ vs. surface-coupling 
enhancement $\kappa$ for the Ising model. The data points $+$, $\times$,
$\Box$, $\bigcirc$, $\bigtriangleup$, $\Diamond$,
and $\ast$  represent
system sizes $L=21, 25, 29, 33, 41, 49$, and $63$, respectively.}
\label{fig03}
\end{figure}
\begin{figure}
\begin{center}
\leavevmode
\epsfxsize 10.0cm
\epsfbox{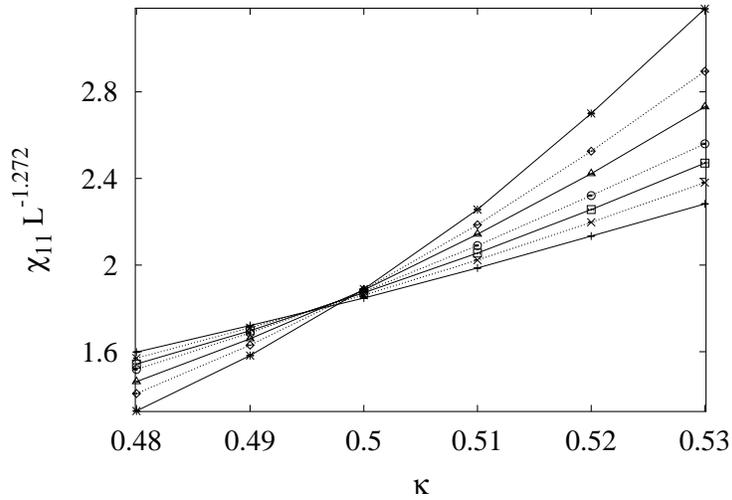}
\end{center}
\caption{Surface susceptibility $\chi_{11}L^{-1.272} $ vs. surface-coupling
enhancement $\kappa$ for the Ising model. The data points $+$, $\times$,
$\Box$, $\bigcirc$, $\bigtriangleup$, $\Diamond$,
and $\ast$  represent
system sizes $L=21, 25, 29, 33, 41, 49$, and $63$, respectively.}
\label{fig031}
\end{figure}

Since it is known that the special transition is located near
$\kappa=(K_1/K)-1 =0.5$, the simulations were performed with surface
enhancements $\kappa$ in the range from $0.46$ to $0.54$, in steps of
$0.01$. The system sizes assumed 18 values in the range
$5\leq L \leq 95$. We sampled several quantities, including the
surface susceptibilities $\chi_{11}$ and $\chi_{12}$, and the universal
ratios $Q_{11}$ and $Q_{12}$.
Part of the data for $Q_{11}$ are shown 
in Fig.~\ref{fig03}, in which the clear intersection indicates the location
$\kappa_{\rm c}^{\rm (s)}$ of the special transition. As mentioned earlier, 
when $\kappa$ deviates from $\kappa_{\rm c}^{\rm (s)}$,
the finite-size behavior of $Q_{11}$ is governed by the surface thermal
exponent $y_{t1}^{\rm (s)}$. We fitted the data for
$Q_{11}$ and $Q_{12}$ by
\begin{eqnarray}
Q_1(\kappa, L)=&& Q_{1{\rm c}}^{\rm (s)} +\sum_{k=1}^4 a_k
(\kappa - \kappa_{{\rm c}}^{\rm (s)} )^kL^{k y_{t1}^{\rm (s)}} +
\sum_{l=1}^4 b_l L^{y_l} + \nonumber \\ &&
c(\kappa -\kappa_{{\rm c}}^{\rm (s)}) L^{y_{t1}^{\rm (s)}+y_{\rm i}} + n
(\kappa -\kappa_{{\rm c}}^{\rm (s)})^2 L^{ y_{t1}^{\rm (s)}} +r_0 L^{y_a}+
\nonumber \\ &&
r_1 (\kappa -\kappa_{{\rm c}}^{\rm (s)}) L^{y_a} + r_2 
(\kappa -\kappa_{{\rm c}}^{\rm (s)})^2
L^{y_a} + r_3 (\kappa -\kappa_{{\rm c}}^{\rm (s)})^3 L^{y_a} \; ,
\label{fitq0}
\end{eqnarray}
where the terms with amplitude $b_l$ account for various finite-size
corrections; and again the subindex $1$ in $Q_1$ and $Q_{1{\rm c}}$
is shorthand for $11$ or $12$, whichever the case may be. The terms with
amplitudes $r_i$ ($i=0,\cdots,3$) are due to the analytic background.
 The derivation of Eq.~(\ref{fitq0})
can be found e.g. in Ref.~\onlinecite{YD2}.
Naturally, we fixed the
exponent $y_1=y_{\rm i}=-0.821 \,(5)$~\cite{YD2},
the exponent of the leading irrelevant scaling field in
the three-dimensional Ising model. In principle,
additional corrections due to irrelevant scaling fields can be induced
by the open surfaces,
so that we set $y_2=y_{i1}$ as an unknown exponent.
In order to reduce the  residual $\chi^2$ without discarding
data for many small system sizes, we included
further finite-size corrections with integer
powers $y_3=-2$ and $y_4=-3$.
The term with coefficient $n$ reflects
the nonlinear dependence of the scaling field on $\kappa$, and the one
with $c$ describes the ``mixed'' effect of the surface
thermal field and the irrelevant field. The terms
with amplitudes $r_0$, $r_1$,
$r_2$, and $r_3$ arise from the analytical part of the free energy,
and the exponent $y_a$ is equal to $2-2y_{h1}^{\rm (s)}$.
As determined later, the surface magnetic exponent at the special
transition is about $y_{h1}^{\rm (s)} = 1.636  \,(1)$, so that
we fixed the exponent $y_a=-1.272$. The fits of $Q_{11}$ yields
$Q_{11c}=0.626 \,(1)$, $\kappa_{\rm c}^{(\rm s)}=0.50214 \,(8)$, and
$y_{t1}^{\rm (s)} =0.7154\,(14)$; from the fit of $Q_{12}$, we obtain
$Q_{12c}=0.2689 \,(1)$, $\kappa_{\rm c}^{(\rm s)}=0.50207 \,(8)$, and
$y_{t1}^{\rm (s)}=0.715\,(4)$.  Next, we simultaneously fitted the
data for $Q_{11}$ and $Q_{12}$ by Eq.~(\ref{fitq0}), and obtain
$\kappa_{\rm c}^{(\rm s)}=0.50208 \,(5)$, and $y_{t1}^{\rm (s)}=0.715\,(1)$.
Our estimate $\kappa_{\rm c}^{(\rm s)}=0.50208 \,(5)$ does not agree well with
the existing results $\kappa_{\rm c}^{(\rm s)}=0.5004 \,(2)$~\cite{CR2,CR1}.
Further, as expected, 
$\kappa_{\rm c}^{(\rm s)}$ does not assume the mean-field value $1/2$.
Attempts to determine the unknown exponent $y_{i1}$ and its associated
amplitude by least-square fitting to the $Q_{11}$ and $Q_{12}$ data
were unsuccessful. These corrections, if present, do not exceed the
detection threshold.  We also fitted the data for the surface
susceptibilities $\chi_{11}$ and $\chi_{12}$ by
\begin{eqnarray}
\chi_1(\kappa, L)=&& L^{2y_{h1}^{\rm (s)}-2} [
a_0 +\sum_{k=1}^4 a_k
(\kappa - \kappa_{{\rm c}}^{\rm (s)} )^kL^{k y_{t1}^{\rm (s)}} +
b_1 L^{y_{\rm i}} +b_2 L^{y_{i1}} +b_3 L^{y_3} + \nonumber \\ &&
b_4 L^{y_4} + c(\kappa -\kappa_{{\rm c}}^{\rm (s)}) L^{y_{t1}^{\rm (s)}+
y_{\rm i}} + n (\kappa -\kappa_{{\rm c}}^{\rm (s)})^2 L^{ y_{t1}^{\rm (s)}}
+r_0 L^{y_a}+ \nonumber \\ &&
r_1 (\kappa -\kappa_{{\rm c}}^{\rm (s)}) L^{y_a} + r_2
(\kappa -\kappa_{{\rm c}}^{\rm (s)})^2 L^{y_a} +
r_3 (\kappa -\kappa_{{\rm c}}^{\rm (s)})^3 L^{y_a} + \nonumber \\ &&
c_{21} (\kappa -\kappa_{{\rm c}}^{\rm (s)}) L^{y_{t1}^{\rm (s)}+y_{i1}}
+ c_{22} (\kappa -\kappa_{{\rm c}}^{\rm (s)})^2
L^{2y_{t1}^{\rm (s)}+y_{i1}} ]\; .
\label{fitchi1}
\end{eqnarray}
Again, the correction exponents
were taken as $y_{\rm i}=-0.821\,(5)$~\cite{YD2},
$y_3=-2$, and $y_4=-3$, and the exponent $y_2=y_{i1}$ was left to be fitted. 
Other than in Eq.~(\ref{fitq0}), we have included in Eq.~(\ref{fitchi1})
the combined effect of the surface thermal field and the
irrelevant field with the unknown exponent $y_{i1}$,
as described by the mixed terms with amplitudes $c_{21}$ and $c_{22}$.
These terms lead to a reduction of the residual $\chi^2$
of the fits, but do not significantly modify the result for the
surface exponent $y_{h1}^{\rm (s)}$.  The surface thermal exponent
was fixed at $y_{t1}^{\rm (s)}=0.715$ as found above. The fit of
$\chi_{11}$ yields
$\kappa_{\rm c}^{\rm (s)}=0.50209 \,(9)$, $y_{h1}^{\rm (s)}=1.636 \,(1)$,
and $y_{i1}=-0.52(2)$. The quoted error margins include the
uncertainty due to the error in $y_{t1}^{\rm (s)}$.
In this case we found clear evidence for corrections, introduced by
the surfaces with an exponent $y_{i1}$.
It is remarkable that such corrections are significant only in combination
with $\kappa$-dependent terms. The data for the surface susceptibility
are shown in Fig.~\ref{fig031} as $\chi_1(\kappa, L) L^{-1.272}$,
where the exponent, which stands for $2-2y_{h1}^{\rm (s)}$, is chosen
such as to suppress the leading $L$-dependence at the special transition. 
As expected, the data display intersections approaching the special
transition as determined above.

\section{$XY$ model}
The bulk critical point of the $XY$ model was determined as 
$K_{\rm c}=0.454\,1655\,(10)$ in Sec. II, which is of sufficient accuracy
to perform the following simulations 
only at this estimated value of $K_{\rm c}$. 

\subsection{Ordinary phase transition}
In analogy with the Ising model, we first let the surface couplings
$K_1$ assume the same values of the bulk couplings, i.e., 
$K_1=K=K_{\rm c}$.
The system size took $14$ values in the range $4 \leq L \leq 48$.
We sampled the surface susceptibilities $\chi_{11}$ and $\chi_{12}$,
and the correlation functions $g_{11}$ and $g_{12}$,
and analyzed the data as we did for the Ising model at the
ordinary phase transition. For instance, the data for $\chi_{11}$ 
and $\chi_{12}$ were also fitted by Eq.~(\ref{fitchi0}), in which
the irrelevant exponent was taken as $y_{\rm i}=-0.789$~\cite{GZJ}. 
The estimates of the surface magnetic 
exponent $y_{h1}^{\rm (o)}$ from various quantities agree;
the result is $y_{h1}^{\rm (o)}=0.781 \,(2)$.

As a consistency test, in analogy with the Ising model, 
we also simulated the surface-enhanced $XY$ model 
as defined in Ref.~\onlinecite{MPN1}, with 
$\epsilon=0.9$ and $0.8$.
As expected, the results for these two cases are in good agreement 
with the above estimate $y_{h1}^{\rm (o)}=0.781 \,(2)$.
However, since the simulations are less extensive in comparison with
those for the case $\epsilon=1$, 
they do not significantly improve the accuracy of $y_{h1}^{\rm (o)}$.

\subsection{Special phase transition}
\begin{figure}
\begin{center}
\leavevmode
\epsfxsize 10.0cm
\epsfbox{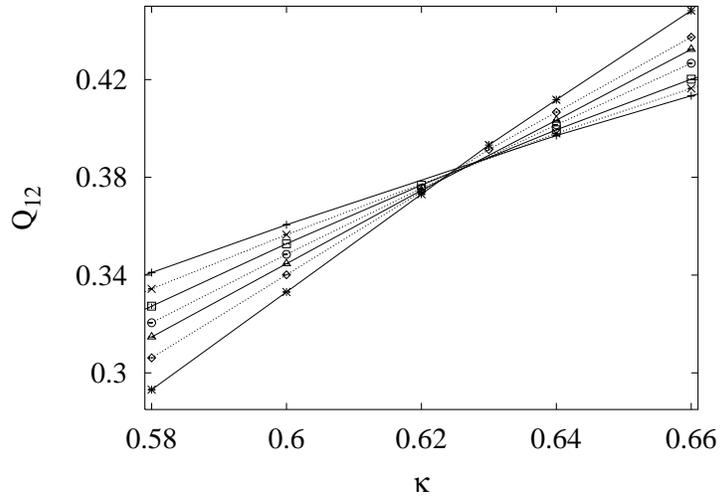}
\end{center}
\caption{Surface dimensionless ratio $Q_{12}$ vs. surface-coupling
enhancement $\kappa$ for the $XY$ model. The data points $+$, $\times$,
$\Box$, $\bigcirc$, $\bigtriangleup$, $\Diamond$,
and $\ast$  represent
system sizes $L=17, 21, 25, 33, 41, 49$, and $63$, respectively.}
\label{fig04}
\end{figure}

As discussed above, the $XY$ model is a marginal case in the sense that
the line of surface phase transitions for  $K<K_{\rm c}$ is 
Kosterlitz-Thouless-like. Still, one would 
expect that, for $K=K_{\rm c}$, the special and 
the extraordinary surface transitions occur.
Therefore, we performed simulations at the estimated bulk critical point 
as given above, and varied the surface enhancement from 
$\kappa=0.48$ to $\kappa=0.68$. The system sizes took on $19$ values
in the range $5 \leq L \leq 95$. The sampled quantities include 
the surface susceptibilities $\chi_{11}$ and $\chi_{12}$, 
the correlation functions  $g_{11}$ and $g_{12}$, and 
the dimensionless ratios  $Q_{11}$ and $Q_{12}$. Part of the data
for  $Q_{12}$ are shown in Fig.~\ref{fig04},
where the intersection clearly indicates that
the special transition occurs near $\kappa_{\rm c}=0.622$. Further,
the increase of the slope of $Q$ as a function of 
finite size $L$ strongly suggests that the surface thermal exponent
at $\kappa_{\rm c}$ is larger than 0, i.e., that the scaling field
associated with $\kappa$ is not marginal at the special transition.
The data for $Q_{11}$ and $Q_{12}$ were fitted by Eq.~(\ref{fitq0}),
in which the leading irrelevant exponent was fixed 
at $y_{\rm i}=-0.789$~\cite{GZJ} and the exponent $y_2=y_{i1}$ was 
left free. We obtain $Q_{11c}=0.840 \,(1)$, $Q_{12c}=0.379 \,(2)$,
$\kappa_{\rm c}= 0.6222 \,(3)$,
and $y_{t1}^{\rm (s)}=0.608 \,(4)$. The fits of $Q_{11}$ and $Q_{12}$ 
do not provide clear evidence for the existence of a term with
exponent $y_{i1}$. 

\begin{table}[!hpt]
\caption{Summary of the results for the surface critical exponents
in the three-dimensional $XY$ and Heisenberg models. MC: Monte Carlo 
simulations, SE: series expansions.}
\label{tab_2}
\begin{center}
\begin{tabular}{||l|l|ll||}
               & {\em ordinary} & \multicolumn{2}{c||}{\em special }  \\
\hline
                      &$y_{h1}$        &$y_{h1}$       &$y_{t1}$    \\
\hline
MC ($XY$)~\cite{DPL1} &$0.74       $   &$--        $   &$--    $    \\
SE ($XY$)~\cite{KO}   &$0.81       $   &$--        $   &$--    $    \\
MC ($XY$)~\cite{MPN1} &$0.790\,(15)$   &$--        $   &$--    $    \\
Present($XY$)         &$0.781\,(2 )$   &$1.675\,(1)$   &$0.608\,(4)$\\
\hline
MC (Heisenberg)~\cite{MPN1}&$0.79\,(2)$& $--    $    & $--    $     \\
Present(Heisenberg)   &$0.813 \,(2 )$  & $--    $    & $--    $   
\end{tabular}
\end{center}
\end{table}
We also fitted the surface susceptibilities  $\chi_{11}$ and
$\chi_{12}$ by Eq.~(\ref{fitchi1}). We obtain
the surface magnetic exponent as $y_{h1}^{\rm (s)}=1.675 \,(1)$.
Further, we find evidence for new finite-size-corrections with
exponent $y_{i1}=-0.44\,(4)$, the major contribution to which comes
form the mixed terms
with amplitudes $c_{21}$ and $c_{22}$ in Eq.~(\ref{fitchi1}).
Results for the surface exponents are summarized in Table~\ref{tab_2}.

\subsection{Extraordinary phase transition}

\begin{table}[!hpt]
\caption{Monte Carlo data for the second moment of surface magnetization
$m_1^2$ and the dimensionless ratio $Q_{11}$ for the three-dimensional
$XY$ model with enhancement $\kappa=1$.}
\label{tab_3}
\begin{center}
\begin{tabular}{||l|lllllll||}
\hline
  $L  $          & $7  $            & $9  $           & $11 $
& $13 $          & $17 $            & $21 $           & $25 $          \\ 
  $m_1^2$        & $0.5653\,(1)$    & $0.5293\,(1)$   & $0.5037\,(1)$
& $0.4839\,(1)$  & $0.4561\,(1)$    & $0.4364\,(1)$   & $0.4216\,(1)$  \\ 
  $Q_{11} $      & $0.96242\,(6)$   & $0.96580\,(6)$  & $0.96878\,(5)$
& $0.97138\,(4)$ & $0.97543\,(3)$   & $0.97835\,(3)$  & $0.98065\,(3)$ \\ 
\hline
  $L  $          & $33 $            & $41 $           & $49 $
& $63 $          & $71 $            & $81 $           & $95 $          \\
  $m_1^2$        & $0.4004\,(1)$    & $0.3859\,(1)$   & $0.3747\,(1)$
& $0.3601\,(1)$  & $0.3540\,(1)$    & $0.3473\,(1)$   & $0.3397\,(1)$  \\
  $Q_{11} $      & $0.98381\,(3)$   & $0.98601\,(3)$  & $0.98748\,(3)$
& $0.98927\,(3)$ & $0.99004\,(3)$   & $0.99085\,(3)$  & $0.99169\,(3)$ \\ 
\hline
\end{tabular}
\end{center}
\end{table}
Two-dimensional surfaces of the $XY$ model do not display spontaneous
long-ranged 
surface order for $K<K_{\rm c}$, but they are in a ferromagnetic state in
the low-temperature region $K>K_{\rm c}$.  Thus the onset of long-range
order on the surface also occurs at $K=K_{\rm c}$.
This differs from the Ising model, where a long-range ordered surface
exists for $K<K_{\rm c}$ if $\kappa>\kappa_{\rm c}$.
We performed simulations at $\kappa=1$ for the critical $XY$ model
with the system sizes 
in the range $7 \leq L \leq 95$. We sampled the second moment of the
surface magnetization $m_1^2$ and the ratio $Q_{11}$;
the data for these two quantities are shown in Table~\ref{tab_3}.

\begin{figure}
\begin{center}
\leavevmode
\epsfxsize 10.0cm
\epsfbox{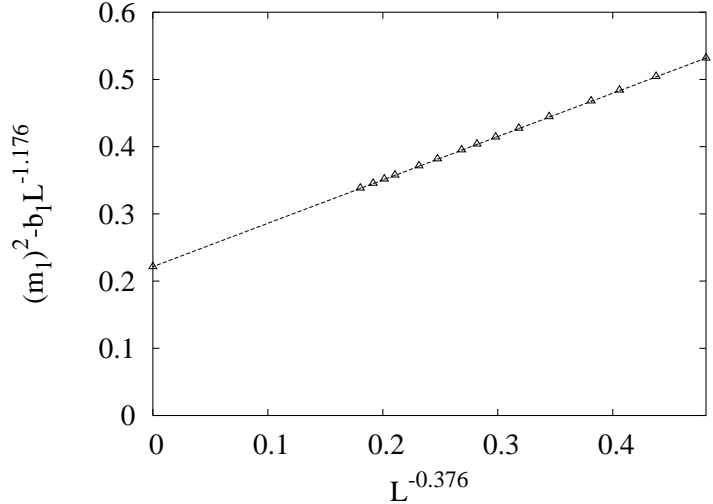}
\end{center}
\caption{Surface magnetization in terms of the quantity
$(m_{1})^2-b_1 L^{-1.2}$ vs. $L^{-2X_{h1}^{\rm (e)}}$
for the $XY$ model at $\kappa=1$, where  the values
$X_{h1}^{\rm (e)}=0.188 \,(5)$ and  $b_1=0.35 \,(5)$ were obtained
from a least-squares fit (see text).}
\label{fig06}
\end{figure}

\begin{figure}
\begin{center} 
\leavevmode
\epsfxsize 10.0cm
\epsfbox{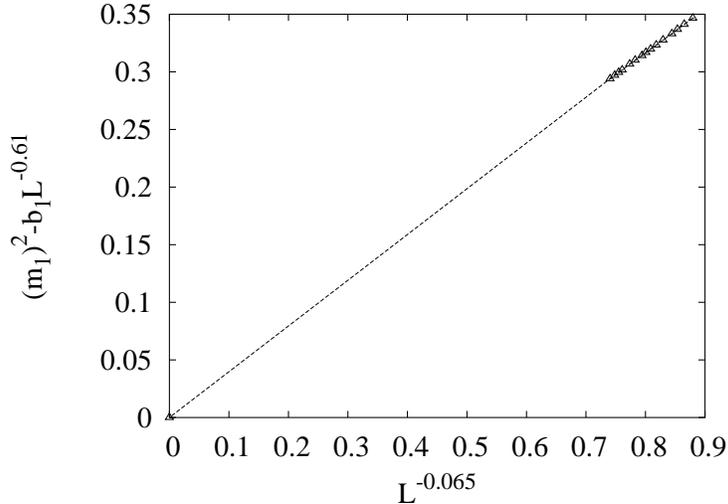}
\end{center}
\caption{Surface magnetization in terms of the quantity
$(m_{1})^2-b_1 L^{-0.61}$ vs. $L^{-0.065}$
for the $XY$ model at $\kappa=1$.} 
\label{fig07}
\end{figure}

In order to analyze the finite-size data in Table~\ref{tab_3}, one 
first requires the proper scaling formulas.
For the extraordinary phase transitions in the $XY$ model, 
there exists some ambiguity, because it is not generally 
clear whether the surfaces undergo a first  or a second order 
transition. 
Nevertheless, in either case, the surfaces
should display some critical singularities, arising from the diverging
bulk correlation length. Thus, we fitted the $m_1^2$ data by
\begin{equation}
m_1^2 (L) =m_{\rm a}^2+ L^{-2X_{h1}^{\rm (e)}} (b_0+b_1 L^{y_1}
+b_2 L^{2 y_1}) \;. 
\label{fitchi2}
\end{equation}
If the transition on the surface is first order at $K=K_{\rm c}$, 
the analytical contribution, $m_{\rm a}^2$, assumes a nonzero value. 
First, we set the exponent $y_1=y_{\rm i}=-0.789$~\cite{GZJ}. 
Satisfactory fits were obtained for all the $m_1^2$ data 
in Table~\ref{tab_3}, with the terms $m_{\rm a}^2$ and those with $b_0$ and 
$b_1$ only. The results are $m_{\rm a} =0.471\,(5)$,
$X_{h1}^{\rm (e)}=0.188\,(5)$, $b_0=0.65\,(1)$, and $b_1=0.35\,(5)$.
The quality of the fit is shown in Fig.~\ref{fig06}.
Further, we fitted the data for the ratio $Q_{11}$ by
\begin{equation}
Q_{11} (L) =
Q_{\rm c}+ b_1 L^{-2X_{h1}^{\rm (e)}} +b_2 L^{-2X_{h1}^{\rm (e)}+y_1}
+b_3 L^{-2X_{h1}^{\rm (e)}+2y_1}+b_4 L^{-2X_{h1}^{\rm (e)}+3y_1} \; ,
\label{fitq1}
\end{equation}
where the irrelevant exponent is fixed at  $y_1=y_{\rm i}=-0.789$~\cite{GZJ}.
The presence of the exponent $X_{h1}^{\rm (e)}$ is due to the
nonzero background contribution $m_a$ in the second moment of the 
magnetization $m_1^2$. We obtain the asymptotic value $Q_{\rm c}=0.9998\,(4)
\approx 1$. From the
results for $m_{\rm a}$ and $Q_{\rm c}$, it seems that the surface transition 
at $K=K_{\rm c}$ and $\kappa=1$ is first order. However, it seems also
possible that the surface magnetization vanishes only very slowly as
the system size $L$ increases, such that the line of extraordinary
transitions on the surfaces is still Kosterlitz-Thouless-like.
Thus, we set $m_{\rm a}$ in Eq.~(\ref{fitchi2}) to zero, and fitted the 
unknown parameters including both $X_{h1}^{\rm (e)}$ and $y_{\rm i}$ to
the $m_1^2$ data. Indeed, we found that our Monte Carlo
data for $m_1^2$ in Table~\ref{tab_3} can be modeled this way, and
we obtain $b_0=0.40\,(1)$, $b_1=0.703\,(6)$,
$X_{h1}^{\rm (e)}=0.0325\,(30)$, and $y_1=-0.545\,(14)$. This fit is
illustrated by Fig.~\ref{fig07}. We also fitted the $Q$ data by
Eq.~(\ref{fitq1}) with $y_1$ fixed at $-0.545$, and the result for
$Q_{\rm c}$ is $Q_{\rm c}=0.9982\,(15)$, which is also consistent with $1$.
In short, our numerical evidence for the surface magnetization of the
three-dimensional $XY$ model is not sufficient to determine whether
the line of transitions for $K=K_{\rm c}$ and $\kappa>\kappa_{\rm c}$ is
first or second order, but settling this matter convincingly would require
extensive simulations, well beyond the scope of the present
investigation.

\section{Heisenberg model}
\begin{figure}
\begin{center}
\leavevmode
\epsfxsize 10.0cm
\epsfbox{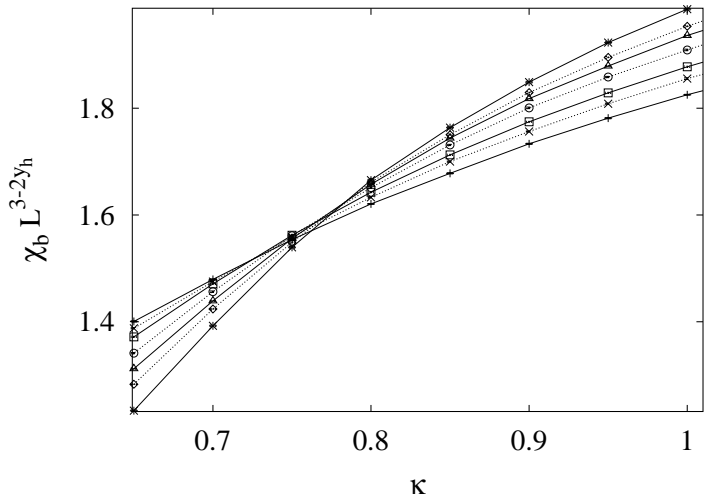}
\end{center}
\caption{Critical bulk susceptibility $\chi_{\rm b}$ of the Heisenberg model
vs. surface enhancement $\kappa$. The data shown along the vertical
axis are scaled with a size-dependent factor $L^{3-2y_h}$ where
$y_h=2.482$ is the bulk magnetic exponent.
The data points $+$, $\times$, $\Box$, $\bigcirc$,
$\bigtriangleup$, $\Diamond$, and $\ast$  represent system sizes
$L=16, 20, 24, 32, 40, 48$, and $64$, respectively.
According to the theory, the scaled susceptibility $\chi_{\rm b} L^{3-2y_h}$
converges with increasing size $L$ to a value that may still depend on 
$\kappa$. The intersections near $\kappa=0.8$ suggest the existence of 
a ``special'' phase transition.
}
\label{fig08}
\end{figure}
\begin{figure}
\begin{center}
\leavevmode
\epsfxsize 10.0cm
\epsfbox{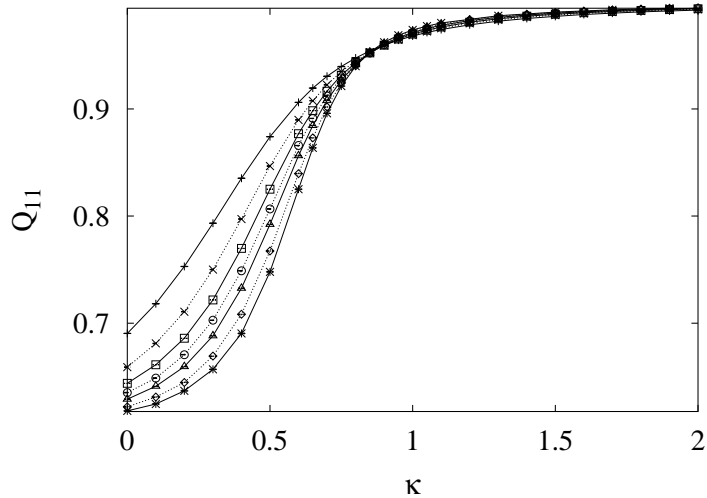}
\end{center}
\caption{Surface dimensionless ratio $Q_{11}$ vs. surface-coupling
enhancement $\kappa$ for the O(3) model. 
The data points $+$, $\times$,
$\Box$, $\bigcirc$, $\bigtriangleup$, $\Diamond$,
and $\ast$  represent
system sizes $L=8, 12, 16, 20, 24, 32$, and $40$, respectively.
For small surface enhancement $\kappa \, \lae \, 0.5$, the ratio $Q_{11}$
converges with increasing $L$ to a nontrivial value near $0.62$, 
just as expected for the ordinary phase transition.  For large
enhancement $\kappa >1$, it seems that the asymptotic value 
$Q_{11}(L \rightarrow \infty)$ is different from 1, and dependent on
$\kappa$. In the intermediate range $0.6 < \kappa <0.9$, the
slope of the $Q_{11}$ data lines increases with $L$. The intersections
of these lines seem to converge to a value near $\kappa=0.8$. This
figure bears much analogy with that for the bulk ratio $Q$ of 
transitions in the Kosterlitz-Thouless universality class.
}
\label{fig09}
\end{figure}
\begin{figure}
\begin{center}
\leavevmode
\epsfxsize 10.0cm
\epsfbox{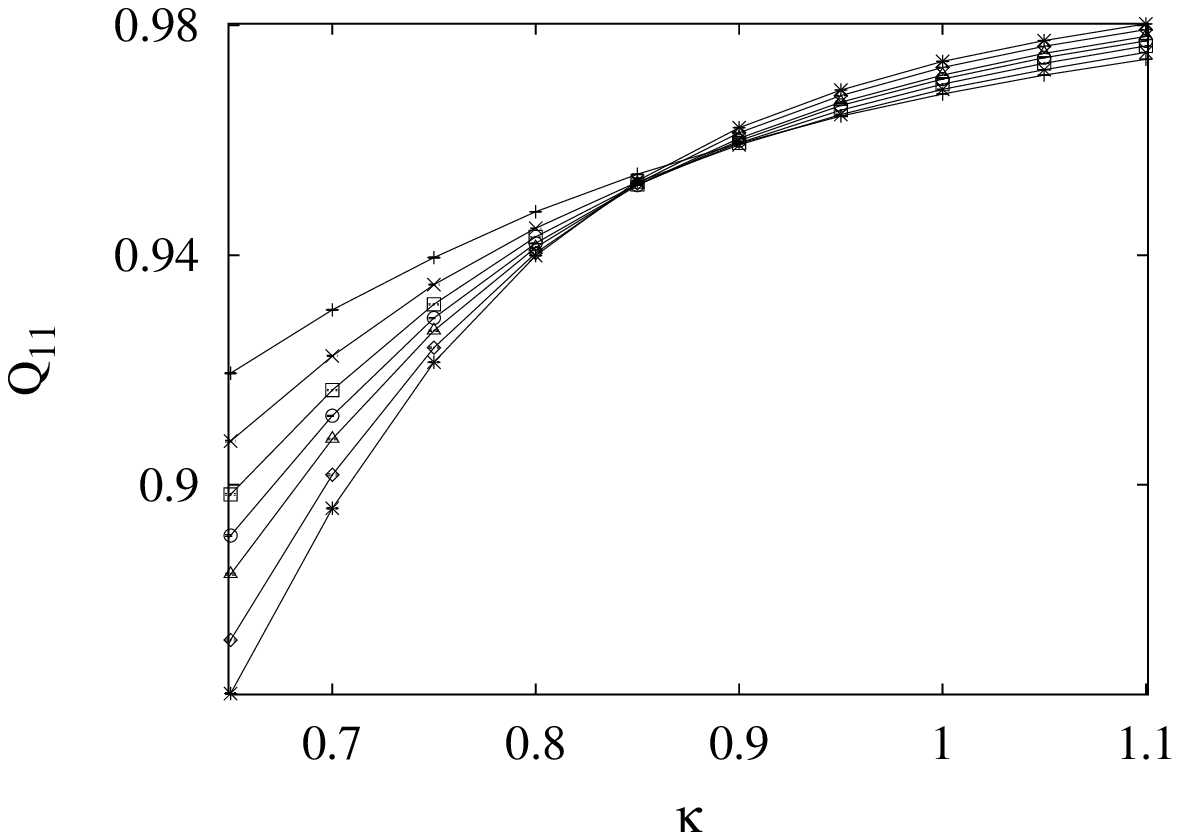}
\end{center}
\caption{Surface ratio $Q_{11}$ in the range $0.65\leq \kappa\leq 1.1$
for the O(3) model. The data points $+$, $\times$, $\Box$, $\bigcirc$,
$\bigtriangleup$, $\Diamond$, and $\ast$  represent
system sizes $L=8, 16, 24, 32, 40, 48$, and $64$, respectively.
The apparent convergence of the intersections of the $Q_{11}$ data
with increasing system size indicates a ``special'' surface transition
near $\kappa=0.80$, in agreement with the results in
Figs.~\ref{fig08} and \ref{fig09}.  }
\label{fig10}
\end{figure}
We simulated the three-dimensional Heisenberg model
at $K_1=K=K_{\rm c}=0.693\,002\,(2)$, as determined in Sec. II.
The system sizes were taken in the range $4 \leq L \leq 64$.
The data for the surface susceptibilities $\chi_{11}$ and $\chi_{12}$,
taken at $\kappa=0$,
were fitted by Eq.~(\ref{fitchi0}). 
Using a similar procedure as that for the $XY$ model, we obtain
$y_{h1}^{\rm (o)}=0.813 \,(2)$ for the ordinary phase transition.
We also determined the bulk susceptibility $\chi_{\rm b}$ and the
dimensionless ratios $Q_{11}$ and $Q_{12}$ for
a range of larger values of the surface enhancement $\kappa$.
The scaled susceptibility $\chi_{\rm b} L^{3-2y_h}$ is shown in
Fig.~\ref{fig08}. The intersections near $\kappa\approx 0.8$ are
very suggestive of a special transition.
The results for $Q_{11}$, shown in Figs.~\ref{fig09} and \ref{fig10},
display similar behavior. For $\kappa \, \lae \, 0.8$, $Q_{11}$ converges
to a  universal constant characteristic of the ordinary transition. 
For $\kappa \, \gae \, 0.8$ the data seem to converge to a
$\kappa$-dependent value. The overall behavior of the results
for $Q_{11}$ resembles that of the ratio $Q$ for bulk transitions
in the Kosterlitz-Thouless universality class, as reported for
the triangular Ising antiferromagnet with nearest- and
next-nearest-neighbor interactions \cite{QB}. An alternative
interpretation would be a special transition with a relevant
exponent $y_{t1}^{\rm (s)}$ only slightly larger than 0.
A convincing numerical
test of the Kosterlitz-Thouless nature of the special transition 
would require simulations beyond the scope of the present work.

\section{Discussion}
We used Monte Carlo techniques and finite-size scaling in order
to obtain new and more accurate results for the bulk 
and surface critical parameters of the three-dimensional Ising, $XY$,
and Heisenberg models. At the ordinary phase transitions, we determined 
the surface magnetic exponents as $y_{h1}^{\rm (o)}(n=1)=0.7374\,(15)$, 
$y_{h1}^{\rm (o)}(n=2)=0.781\,(2)$, and $y_{h1}^{\rm (o)}(n=3)=0.813\,(2)$. 
These values are in a satisfactory agreement with earlier results
\cite{MPN1}, namely $y_{h1}^{\rm (o)}(n=1)=0.740\,(15)$, 
$y_{h1}^{\rm (o)}(n=2)=0.790\,(15)$, and $y_{h1}^{\rm (o)}(n=3)=0.79\,(2)$,
as shown in Table~\ref{tab_2}.
Since the bulk thermal exponent $y_t$ of the O$(n)$ model 
decreases with increasing $n$, these results suggest that the
surface exponent $y_{h1}^{\rm (o)}$ is a decreasing function of $y_t$.
The same seems to hold true for the two- and three-dimensional
Potts models, as may be concluded on the basis of the
following evidence. In three dimensions, the surface magnetic exponent
for the $q \rightarrow 0$ and $q \rightarrow 1$ Potts models are
$y_{h1}^{\rm (o)}=2$ and $1.0246 \,(6)$~\cite{YD5}, respectively.
The former model is generally referred to as
the uniform spanning tree~\cite{FYW}, while 
the $q \rightarrow 1$ Potts model
reduces to the bond percolation model. For the two-dimensional Potts
model, from the conformal field theory, the exponent $y_{h1}^{\rm (o)}$ 
is exactly known as $y_{h1}^{\rm (o)}=2-3/(3-y_t)$~\cite{JLC1}, 
which is a decreasing function of  the bulk thermal exponent $y_t$.
Further, if one applies the above expression to the tricritical branch
of the Potts model in two dimensions, one obtains that the surface
magnetic scaling field is irrelevant at the ordinary phase transition.
Starting from this observation, it was  found~\cite{YD6} that rich
surface phase transitions can also occur in some two-dimensional
systems, although their ``surfaces'' are just one-dimensional edges.

In the present work, we also located the special transitions of
the Ising and the $XY$ model on the simple-cubic lattice, and obtained
numerical estimates of the corresponding renormalization exponents.
While the surface transition of the three-dimensional $XY$ model
is Kosterlitz-Thouless-like, and the line of surface transitions 
connects to the special transition point, our numerical data did not
yield evidence for corrections to scaling due to a marginal field
at the special transition. 

Finally, we note that the surface-critical
behavior of the O$(1)$, O$(2)$ and O$(3)$ models is rather dissimilar for
large surface enhancements. For the O$(1)$ model, spontaneous surface
order exists even below the bulk critical coupling $K_{\rm c}$; for the
O$(2)$ model it exists for $K>K_{\rm c}$ and possibly for $K=K_{\rm c}$;
and for the O$(3)$ model only for $K>K_{\rm c}$.
In line with the bulk critical singularity, the O$(n)$ surface
critical behavior is thus seen to become less singular with
increasing $n$. This is also evident from our analyses of the special
transitions, which yield relevant exponents $y_{t1}^{\rm (s)}$ for
the O(1) and O(2) models but allow a marginal exponent for the
O(3) model. Since the lower critical dimensionality of the special
transition \cite{KB1} is 3 for $n>2$, it seems plausible that the
range  $\kappa>\kappa_{\rm c} $ corresponds with a line of fixed points and 
$\kappa$-dependent critical surface exponents, in agreement with 
an analysis of the surface magnetization by Krech \cite{MK}.
Indeed, the data in Figs.~\ref{fig08} and \ref{fig09}
are suggestive of a
Kosterlitz-Thouless-like scenario involving a nonuniversal range
of $Q$-values such as found earlier in the different context of the
Ising triangular antiferromagnet \cite{QB}.
\begin{acknowledgments}
The authors are indebted to Dr. J.R. Heringa and
X.F. Qian for valuable discussions.
This research was supported by the Dutch FOM foundation (``Stichting
voor Fundamenteel Onderzoek der Materie'') which is financially supported by
the NWO (``Nederlandse Organisatie voor Wetenschappelijk Onderzoek'').
This research was supported in part by the United States National
Science Foundation under grant number ITR 0218858.
\end{acknowledgments}

\end{document}